\documentclass[authoryear,preprint,12pt,authoryear]{elsarticle}

\usepackage{amssymb}
\usepackage{amsmath}
\usepackage{graphicx}
\usepackage{caption}
\usepackage{subcaption}
\usepackage{booktabs,siunitx}
\usepackage[protrusion=true,expansion=true]{microtype} 
\usepackage[utf8]{inputenc}
\usepackage{bbm}
\usepackage{multirow}

\def\doubleunderline#1{\underline{\underline{#1}}}

 \usepackage{color}
 \usepackage{xcolor}
\usepackage{verbatim}

\journal{ArXiv.org, }

\begin{document}

\begin{frontmatter}



\title{Modelling a conductive-capacitive medium using the boundary element method}


\author{Bram Schoonjans*, Johan Deconinck}

\address{Vrije Universiteit Brussel, Faculteit Ingenieurswetenschappen, Elektrotechniek en Energietechniek, Pleinlaan 2, 1050 Brussel, Belgium, *e-mail address: bram.schoonjans@vub.ac.be}

\begin{abstract}
In this paper a generalized fundamental solution using the boundary element method to solve the Helmholtz equation is proposed. It is observed that the commonly used fundamental solution is only valid for good conductors since the capacitive effect of the considered medium is always neglected.
\\ \\
By the use of the well-known Lorentz gauge condition a fundamental solution which incorporates the phase as well as the attenuation transmission coefficients is derived by the authors. Next, a model is developed using this proposed fundamental solution for modelling a coating layer of a buried pipeline. Subsequently, a model of a buried coated pipeline in close proximity to a high voltage power line is developed and numerically implemented.
\\ \\
Finally, the model is used to simulate two configurations in order the verify the proposed general fundamental solution of the boundary element method for electromagnetic field problems. Hereby, its validity is proven and it is shown that the generalized solution and related model can be used for industrial applications.
\end{abstract}


\end{frontmatter}



\section*{Introduction}

In order to transport natural gas or oil, steel pipes are used. These pipes are constructed in order to supply resources at different locations located at great distances apart from each other. To avoid visual inconvenience, these pipelines are often buried at specific locations.
\\ \\
Besides these pipelines, there is also the necessity to transport electrical energy over long distances and to supply at the same locations. Due to the scarcity of land in the developed world, there is the tendency to allocate the same utility to these means of energy transport.
\\ \\
As a result, pipelines and power lines come into close proximity and an electromagnetic coupling occurs between both. The electromagnetic field originating from the power line induces a voltage on the buried pipeline which leads to risks for human safety and structural damage.
\\ \\
Since the calculation of the level of induced voltages is dependent of a multitude of parameters, it is obvious an analytical solution of a real life situation is excluded. Therefore, there is the need to model the problem using numerical calculations methods.
\\ \\
In 1926, Carson [\cite{Carson}] was the first who modelled electromagnetic field problems involving overhead powerlines. Using equivalent circuits this lead to his well-known formulae. In 1949 Sunde extended the model towards a multilayered earth [\cite{Sunde}].
\\ \\
Later on simulation software was developed using a transmission line model (TLM), like ECCAPP [\cite{Dawalibi,Dawalibi2}], that was further improved by Bortels in [\cite{Bortels}].
\\ \\
Munteanu [\cite{Munteanu}] however didn't use a TLM to tackle to model, but developed a method that combined a finite element method (FEM) with a boundary element method (BEM). This model consists of a one dimensional discretization of the pipeconductor using FEM and whereby the surrounding earth is modelled using a three dimensional BEM. In fact this was an extension of the DC model of Brichau [\cite{Brichau}] to AC.
\\ \\
However, it was observed that a hiatus was present in the model, namely the way of dealing with unbounded regions, in this case the soil surrounding the pipeline. In such a situation, the BEM requires a specific solving method, namely to calculate the geometry dependent coefficient $c_i$ in the general expression in an indirect way. This problem was solved by the authors through the proposal of an innovative yet general method based on the skin depth [\cite{Schoonjans}].
\\ \\
Next, without anyhow devaluating the excellent work performed by Munteanu, another limitation of his model is the absence of a coating layer. Obviously, this is a drawback when real life situations need to be investigated. Since a coating layer cannot be perceived as a good conductor and because the fundamental solutions described in the literature are only valid when modelling conductive media, a more general expression of the fundamental solution for the BEM when modelling high-voltage low frequency electromagnetic field problems is derived in this paper. This general expression, which can be approximated for either good conductors or for dielectric materials, is used to model the coating layer.    
\\ \\
In the first part a briefly overview is given of electromagnetic waves in order to define the attenuation as well as the phase coefficients. In the second part these coefficients are used to derive the most general fundamental solution for the BEM needed to model the coating layer. In the following part the model is discussed and in the final part of this paper simulation results are discussed and compared with those of a TLM.
\section{Description of the electromagnetic waves}
\subsection{Wave equations}
In homogeneous, linear and isotropic media, electromagnetic waves can be described in a direct way by the electric field $\vec{E}$ and the magnetic field $\vec{B}$ [\cite{Griffiths}]
\begin{subequations}
\begin{equation}
\Delta\vec{E}-\mu\sigma\frac{\partial\vec{E}}{\partial t}-\mu\epsilon\frac{\partial^2\vec{E}}{\partial t^2}=0,
\end{equation}
\begin{equation}
\Delta\vec{B}-\mu\sigma\frac{\partial\vec{B}}{\partial t}-\mu\epsilon\frac{\partial^2\vec{B}}{\partial t^2}=0,
\end{equation}
\label{eq:1}
\end{subequations}
where $\mu$, $\sigma$ and $\epsilon$ represents respectively the permeability, conductivity and permittivity of the medium wherein the wave propagates. As generally known, equations (\ref{eq:2}a) and (\ref{eq:2}b) admit plane-wave solutions that are expressed by [\cite{Reitz}]
\begin{subequations}
\begin{equation}
\vec{E}(\vec{r},t)=\tilde{E}_0 e^{-i(\tilde{k}\vec{r}-\omega t)},
\end{equation}
\begin{equation}
\vec{B}(\vec{r},t)=\tilde{B}_0 e^{-i(\tilde{k}\vec{r}-\omega t)},
\end{equation}
\label{eq:2}
\end{subequations}
in which $\tilde{E}_0$ and $\tilde{B}_0$ are functions of the applied boundary conditions and $\tilde{k}$ is the complex wave number. The complex wave number equals
\begin{equation}
\tilde{k}=\alpha+i\beta,
\end{equation}
\begin{subequations}
where the attenuation transmission coefficient $\alpha$ and the phase transmission coefficient $\beta$ are [\cite{Stratton}]
\begin{equation}
\alpha\equiv\omega\left [\frac{\mu\epsilon}{2}\left ( \sqrt{1+\left(\frac{\sigma}{\epsilon\omega}\right)^2}+1\right ) \right ]^{1/2},
\end{equation}
\begin{equation}
\beta\equiv\omega\left [\frac{\mu\epsilon}{2}\left ( \sqrt{1+\left(\frac{\sigma}{\epsilon\omega}\right)^2}-1\right ) \right ]^{1/2}.
\end{equation}
\label{eq:ab}
\end{subequations}
\subsection{Potential functions}
In numerical electromagnetics it is more common to use potential functions in order to execute the computations rather than to calculate the electric field $\vec{E}$ and the magnetic field $\vec{B}$ in a direct way.
\\
\\
Therefore the magnetic vector potential $\vec{A}$ and the electric scalar potential $\Phi$ are introduced and defined as
\begin{subequations}
\begin{equation}
\vec{B}=\nabla\times\vec{A},
\end{equation}
\begin{equation}
\vec{E}=-\nabla\Phi-\frac{\partial \vec{A}}{\partial t}.
\label{veldE}
\end{equation}
\end{subequations}
Field equations are invariant to gauge transformations, allowing prescribing any convenient gauge condition. The aim is to model a coating layer of a buried pipeline. For this medium, the capacitive characteristics cannot be neglected in comparison to the conductive properties. Therefore, the Lorentz gauge condition used in the following form
\begin{equation}
\nabla\cdot\vec{A}=-\mu\left(\sigma+i\omega\epsilon\right)\Phi.
\label{gauge}
\end{equation}
Using this condition, Maxwell's equations can be reduced to the following Helmholtz equations that describe electromagnetic field problems without loss of generality,
\begin{subequations}
\begin{equation}
\Delta\vec{A}-\lambda^2\vec{A}=0,
\end{equation}
\begin{equation}
\Delta	\Phi-\lambda^2\Phi=0.
\end{equation}
\label{eq:helm}
\end{subequations}
The wavenumber equals
\begin{equation}
\lambda=\sqrt{-i\omega\mu\left(\sigma+i\omega\epsilon\right)}
\end{equation}
and characterizes the medium where the Helmholtz equation holds.
\section{Fundamental solutions in the BEM}
If a Helmholtz equation of an arbitrary function $f$ in a region $\Omega$ with boundary $\Gamma$ as a governing equation is considered, the BEM can be used to calculate any function value of any arbitrary point $p_i$ within the region by use of the following general expression of the method
\begin{equation}
c_i f(p_i)+\int_\Gamma f\frac{\partial \mathcal{G}}{\partial n}d\Gamma =\int_\Gamma \frac{\partial f}{\partial n}\mathcal{G}d\Gamma,
\label{eq:bem1}
\end{equation}
with $\mathcal{G}$ the so-called fundamental solution or Green function of the governing equation and $n$ the unit outward normal to the boundary $\Gamma$ [\cite{Brebbia}].
\\ \\
Generally, the fundamental solution listed in BEM reference books like [\cite{Brebbia,Shen,Sadiku,Sykulski}] related to electromagnetic field problems neglects the capacitive effect of the modelled medium. In order to take this effect into account, the fundamental solution takes the form
\begin{equation}
\mathcal{G}_1=\frac{e^{-\beta |\vec{r}| \left(1+i\frac{\alpha}{\beta}\right)}}{4\pi |\vec{r}|},
\label{fundsol1}
\end{equation}
with again $\alpha$ and $\beta$ the coefficients in equations (\ref{eq:ab}a-b).
\\ \\
Note that if the capacitive effect is neglected - i.e. a good conductor is considered - it is stated that [\cite{Balanis}]
\begin{equation}
\left(\frac{\sigma_p}{\omega\epsilon_p}\right)^2\gg 1,
\label{g_cond}
\end{equation}
thus the attenuation as well as the phase transmission coefficients converges to the following value [\cite{Balanis}]
\begin{equation}
\alpha\simeq\beta\simeq\sqrt{\frac{\omega\mu\sigma}{2}},
\label{good_cond}
\end{equation}
which leads to the well-known fundamental solution for good conductors,
\begin{equation}
\mathcal{G}_2=\frac{e^{-\beta |\vec{r}|\left(1+i\right)}}{4\pi |\vec{r}|}.
\label{fundsol2}
\end{equation}
If dielectric materials are considered the attenuation transmission coefficient converges to [\cite{Balanis}],
\begin{equation}
\alpha\simeq\omega\sqrt{\mu\epsilon},
\label{notgood_cond1}
\end{equation}
and the phase transmission coefficient to [\cite{Balanis}],
\begin{equation}
\beta\simeq\frac{\sigma}{2}\sqrt{\frac{\mu}{\epsilon}}.
\label{notgood_cond2}
\end{equation}
The fundamental solution for a dielectric material is therefore,
\begin{equation}
\mathcal{G}_3=\frac{e^{-|\vec{r}|\left(\frac{1}{2}\sqrt{\frac{\mu\sigma}{\epsilon}}+i\omega\sqrt{\epsilon\mu}\right)}}{4\pi |\vec{r}|}.
\label{fundsol3}
\end{equation}
\section{Combined BEM-FEM-model for coated buried pipelines}
\begin{figure}[h]
	\centering
		\includegraphics[width=70mm]{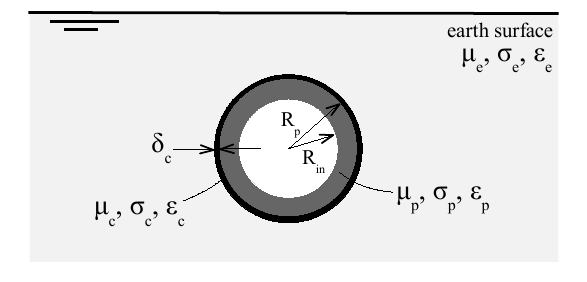}
	\caption{Electromagnetic characterization of the three media.}
\label{fig:1}
\end{figure}

\begin{figure}[h]
	\centering
		\includegraphics[width=50mm]{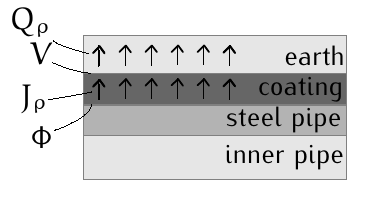}
	\caption{Voltages and current densities in the different media.}
\label{fig:2}
\end{figure}

In what follows the model described by [\cite{Munteanu}] is completed by incorporating the coating layer between the pipeline and earth that is always present.
\\ \\
The buried pipeline is modelled using the FEM, while the coating layer is modelled using the BEM by making use of the general fundamental solution in (\ref{fundsol1}).
\\ \\
The surrounding earth is modelled using the BEM whereby the innovative method as described in [\cite{Schoonjans}] is implemented.
\\ \\
Each medium is, in general, characterized by its corresponding permeability $\mu$, conductivity $\sigma$ and permittivity $\epsilon$, see figure \ref{fig:1}. However, for the steel pipe as well as for the surrounding soil $\epsilon_p$ respectively $\epsilon_s$ can be neglected.
\\ \\
Due to the inductive coupling between buried pipelines and neighbouring high voltage AC power lines, a voltage $\Phi$ is induced inside the steel medium. Because of the high conductivity of this medium it is assumed that $\Phi$ is radial independent, meaning that the potential on the inside of the coating layer over the whole boundary equals $\Phi$.
\\ \\
Through the coating layer a radial current density $J_\rho$ will flow because of the voltage difference between the potential $\Phi$ in the steel pipe and the potential $V$ of the surrounding earth.
\\ \\
The potential on the outside of the coating layer is defined as $V$ and a current density from the coating layer to the surrounding earth will flow as well and is called $Q_\rho$.
\\
\\
These different physical quantities are schematically presented in figure \ref{fig:2}.
\\
\\
\subsection{The steel pipe modelled using the FEM}
In the steel pipe the material properties are noted with the subscript $p$. Since steel pipes are good conductors, condition (\ref{g_cond}) holds and the wavenumber in (\ref{eq:helm}b) reduces to $\lambda=\sqrt{-i\omega\mu_p\sigma_p}$ and the Helmholtz equation that holds in the steel pipe is
\begin{equation}
\Delta\Phi-i\omega\mu_p\sigma_p\Phi=0.
\label{fem_helm}
\end{equation}
This so-called \textit{internal} Helmholtz equation is solved using an axisymmetric coordinate system that is always aligned with the centre of the pipeline.
\\
\\
Since the dimensions in axial z-direction are orders of magnitude higher compared to those in the radial $\theta$-direction any variance of the potential in the $\theta$-direction is neglected.
\\
\\
The Helmholtz equation (\ref{fem_helm}) in an axiymmetric coordinate system and neglecting the $\theta$-direction is
\begin{equation}
\frac{\partial^2 \Phi}{\partial z^2}+\frac{1}{\rho}\frac{\partial}{\partial\rho}\left ( \rho\frac{\Phi}{\partial\rho} \right )=i\omega\mu_p\sigma_p\Phi.
\label{helm_axi}
\end{equation}
In the gauge condition (\ref{gauge}) the capacitive effect is neglected and substituted into (\ref{helm_axi}). The magnetic vector potential $\vec{A}$ originates from the high voltage power line - see section \ref{mvp} - and since the wall thickness of the pipe is negligible compared to its distance to the power line any variation of the radial component of $\vec{A}$ in that direction will also not be taken into account.
\\
\\
Equation (\ref{helm_axi}) becomes
\begin{equation}
\frac{\partial^2 \Phi}{\partial z^2}+\frac{1}{\rho}\frac{\partial}{\partial\rho}\left ( \rho\frac{\Phi}{\partial\rho} \right )=-i\omega\frac{\partial A_z}{\partial z}.
\end{equation}
\\
The latter equation is reduced to an equation along only the $z$-direction by integrating it over the pipe surface,
\begin{equation}
\int_{0}^{2\pi}\int_{R_{in}}^{R_{p}}\sigma_p\left (\frac{\partial^2 \Phi}{\partial z^2}+\frac{1}{\rho}\frac{\partial}{\partial\rho}\left ( \rho\frac{\Phi}{\partial\rho} \right )+i\omega\frac{\partial A_z}{\partial z} \right )\rho d\theta d\rho=0,
\end{equation}
\\
and assuming that no radial current flows into the inner side of the pipeline, the governing equation for the inner problem becomes
\begin{equation}
\frac{\partial^2\Phi}{\partial z^2}-\frac{2\pi R_p}{\sigma_p S_p} J_\rho+i\omega\frac{\partial A_z}{\partial z}=0,
\label{eq17}
\end{equation}
\\
wherein $J_\rho=\sigma_p\frac{\partial\Phi}{\partial\rho}|_{R_p}$ becomes a source term.
\\ \\
The standard FEM with linear elements is used to discretize (\ref{eq17}) yielding for each element of length $\ell$ with nodes $i$ and $j$ the following set of equations:
\begin{equation}
\begin{cases}
\begin{split}
I_i  & = (\Phi_i-\Phi_j)\cdot S_{P} + k\cdot\left ( \frac{J_{\rho_i}}{3}+\frac{J_{\rho_j}}{6}\right) \\ & -j\omega \ell \cdot S_{P}\cdot\left( \frac{A_{z_i}+A_{z_j}}{2}\right ), \\
- I_j  & = (\Phi_j-\Phi_i)\cdot S_{P} + k\cdot\left ( \frac{J_{\rho_i}}{6}+\frac{J_{\rho_j}}{3}\right) \\ & +j\omega \ell \cdot S_{P}\cdot\left( \frac{A_{z_i}+A_{z_j}}{2}\right ),
\end{split}
\end{cases}
\label{eq:fem2}
\end{equation}
with $I$ the current in the axial direction, $k=2\pi R_p\ell$ the surface of revolution per element and $S_{P}=\sigma_p S_p / \ell$ the axial conductivity.
\\ \\
In this approach it is assumed that $J_\rho$ varies linearly within each element between the values $J_{\rho_i}$ and $J_{\rho_j}$. The same reasoning holds for $A_z$.
\\ \\
Writing these two equations for all elements yields the system of equations representing the internal potential problem in the pipes noted with superscript $p$ using the FEM:
\begin{align}
\label{tsm1}
\begin{bmatrix}
\doubleunderline{S}^p & \doubleunderline{K}^p.
\end{bmatrix}
\left \{
\begin{array}{c} \underline{\Phi} \\  \underline{J}_\rho \end{array} \right \}
=
\begin{bmatrix}
 \doubleunderline{A}^{p}
\end{bmatrix}
\left \{
\begin{array}{c}  \underline{A}_z \end{array} \right \}.
\end{align}
In this system of equations the axial component of $\vec{A}$ is an imposed source term but it is obvious that since two unknowns remain, this system of equations still needs to be extended.
\subsection{Modelling the coating layer and the earth using the BEM}
Two boundaries are distinguished, namely that of the steel pipe and that of the coating layer. The coating layer, bounded by the steel pipe and surrounding earth, as well as the surrounding earth, are modelled using the BEM.
\subsubsection{The coating layer}
\begin{figure}[h]
	\centering
		\includegraphics[width=100mm]{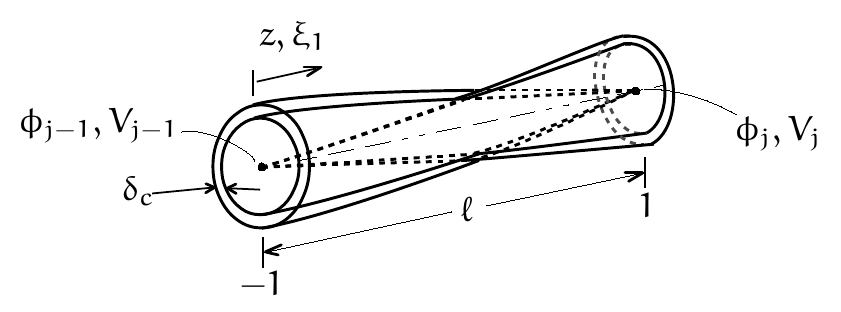}
	\caption{Modelling the coating layer using boundary elements.}
\label{fig_coat_bem}
\end{figure}

A purpose of the coating layer is to protect the pipeline against corrosion. Hence, it is obvious that a coating layer has a very high electrical resistance compared to the steel pipeline and the surrounding earth. Thus, one could argue to introduce the coating layer in the model as an impedance with a high ohmic resistance between the soil and the steel pipe. This approach however does not work properly since it only takes into account the voltage drop over the coating layer seen at each node. This was observed after numerous calculations.
\\ \\
Note that the strength of the BEM is that the influence of each neighbouring node is taken into account. Therefore the coating layer is modelled using the BEM. A solution with the well-known fundamental solution (\ref{fundsol2}) with a high resistance leads to numerical errors. Again, also this observation was made after numerous computations. Thus, to be able to use the BEM for modelling conductive-capacitive media, the proposed general expression of the fundamental solution (\ref{fundsol1}) needs to be used. This fundamental solution may be approximated when dealing with dielectric materials, see (\ref{fundsol3}), yet due to the specific characteristics of a coating layer, it  cannot be perceived as a dielectric material for the following reasons.
\\ \\
In general, a coating layer is, from an electromagnetic point of view, characterized by its permeability $\mu$, conductivity $\sigma$ and permittivity $\epsilon$. These characteristics combined with the angular frequency $\omega$ can be used to calculate the attenuation transmission coefficient $\alpha$ and the phase transmission coefficient $\beta$, see (\ref{eq:ab}a) respectively (\ref{eq:ab}b). For a polyethylene coating layer with the characteristics listed in Table \ref{chap10_tab1}, the attenuation transmission coefficient $\alpha$ and the phase transmission coefficient $\beta$ becomes $2.36957\times10^{-6}$ Np/m respectively $3.33211\times10^{-6}$ rad/m. When on the one hand the approximation formula for good conductors is used, see (\ref{good_cond}), a relative error of $18.6\%$ respectively $15.7\%$ is made. On the other hand, when the approximation formulas for good dielectrics are used, see (\ref{notgood_cond1}) and (\ref{notgood_cond2}), the relative errors are $42.2\%$ respectively $29.7\%$.
\begin{table}[h]
\caption{Characteristics of polyethylene coating layer [\cite{CIGRE}].}
\begin{center}
\begin{tabular}{lccr}
\hline
Thickness of the coating $\delta_c$ &\ &\ & $4$mm\\
Relative permittivity of the coating $\epsilon_r$ &\ &\ & $5$\\
Relative permeability of the coating $\mu_r$ &\ &\ & $5$\\
Resistance of the coating $\rho_c$ &\ &\ & $25\times10^6$ $\Omega$m\\
Angular frequency $\omega$ &\ &\ & $100\pi $ rad/s\\
\hline \hline
\end{tabular}
\end{center}
\label{chap10_tab1}
\end{table}
\\ \\
Thus, the coating layer is modelled by use of the most general form of the fundamental solution $\mathcal{G}_1$, see (\ref{fundsol1}). The subscript $c$ is used to distinguish the material properties. The coating layer around the steel pipe is discretized in the $z$ direction in elements with length $\ell$ that coincide with the elements used to discretize the steel pipe. It is assumed that the current density $J_\rho$ that enters or leaves the coating at the steel surface as well as the current density $Q_\rho$ that enters or leaves the coating at the earth surface are only varying along the $z$-direction and in a linear dependency of $z$. These assumptions are complying with linear pipe elements, see [\cite{Munteanu}], and the FEM discretization of the pipe. Further, for the potentials $\Phi$ and $V$ the same assumptions are made. Using the BEM, one has to integrate in each element over the outer surfaces of the coating: once at the outer radius of the pipeline, and once at the outer radius plus the coating thickness $\delta_c$, see figure \ref{fig_coat_bem}.
\\
\\
As a result, if $N$ elements  are implemented to model the coating layer, the following BEM matrix system of dimension $2N\times 2N$ is obtained
\begin{align}
\begin{bmatrix}
\label{echap10_eq2}
\doubleunderline{H} ^c
\end{bmatrix}
\left \{
\begin{array}{c} \underline{\Phi} \\ \underline{V}  \end{array} \right \}
=
-
\begin{bmatrix}
 \doubleunderline{G}^c / (\sigma_c+i\omega\epsilon_c)
\end{bmatrix}
\left \{
\begin{array}{c}  \underline{J}_\rho \\ \underline{Q}_\rho \end{array} \right \},
\end{align}
\\
with $\doubleunderline{H} ^c$ and $\doubleunderline{G} ^c$ being the classical but complex BEM $H$ and $G$ matrices for the coating, hence the superscript $c$.
\\ \\
Furthermore, it is obvious that the coating layer is a closed system: the incoming current density summed over the whole inner surface equals the total outcoming current density over the exterior one, or
\begin{equation}
\sum_{elements}\int_{S_{el_{int}}} J_{\rho_i}dS+\sum_{elements}\int_{S_{el_{ext}}} Q_{\rho_i}dS=0.
\end{equation}
with again $N$ the number of elements. Observe that this is not necessarily true within each element.
\\ \\ 
As the coating is a closed system, the diagonal terms of the $H^c$-matrix may be calculated as follows,
\\
\begin{equation}
H^{c}_{ii}=-\sum\limits_{j=1}^{2N} H^{c}_{ij}.
\end{equation}
\subsubsection{The surrounding earth}
The sub- and superscript $s$ annotates the material properties of the soil. In general the capacitive effect of the soil surrounding the buried pipeline is not taken into account since it can be considered to be a good conductor [\cite{Gary}]. Therefore the fundamental solution, $\mathcal{G}_2$, see (\ref{fundsol2}), is used. 
\\
\\
The function $f$ in the general expression of the BEM, see (\ref{eq:bem1}), is now the potential $V$ on the outside of the coating layer. Combined with (\ref{veldE}), the Lorentz gauge condition for good conductors and the constitutive equation of conductivity  [\cite{Munteanu}], one has,
\begin{equation}
c_i V_i+  \int_\Gamma V \frac{\partial \mathcal{G}_2}{\partial n}d\Gamma =\int_\Gamma \left(\frac{Q_\rho}{\sigma_{s}}+i\omega A_\rho\right)\mathcal{G}_2 d\Gamma,
\end{equation}
with $\sigma_{s}$ the conductivity of the surrounding soil and $A_\rho$ the radial component of the magnetic vector potential. The equation describing the influence of the surrounding earth becomes\\
\begin{align}
\begin{bmatrix}
\label{chap10_eq7}
\doubleunderline{H}^{s} & -\doubleunderline{G}^{s} / \sigma_{s}
\end{bmatrix}
\left \{
\begin{array}{c} \underline{V} \\ \underline{Q}_\rho  \end{array} \right \}
=
\begin{bmatrix}
i \omega \doubleunderline{G}^{s}
\end{bmatrix}
\left \{
\begin{array}{c}  \underline{A}_{\rho} \end{array} \right \}.
\end{align}
\\
The effect of the discontinuity between the air and the ground is taken into account using the principle of the mirror technique as discussed in [\cite{Munteanu}].
\\ \\
Following the same reasoning as in [\cite{Schoonjans}] the diagonal terms in the latter equation are calculated as follows
\begin{equation}
H^{s}_{ii}=-\sum\limits_{j=1}^{N} H^{s}_{ij}+e^{-2\pi}\left(1+2\pi\left(1+i\right)\right ).
\label{eq24}
\end{equation}
\subsection{The magnetic vector potential}
\label{mvp}
The applied source conditions in equations (\ref{tsm1}), (20) and (23) are the $z$ and $\rho$ components of the magnetic vector potential $\vec{A}$ originating from the high voltage power line. The general formula to compute $\vec{A}$ is [\cite{Shen}]
\begin{equation}
\Delta\vec{A}-i\omega\mu\sigma\vec{A}=-\mu\vec{J}_s,
\label{veca1}
\end{equation}
whereby $\vec{J}_s$ is the current density of one phase of the power line.
\\ \\
Since the current source is located in the air and the observation point is in the soil, the modified image theory needs to be applied, see [\cite{Grcev}].
\\ \\
In the case of power line frequencies (50 and 60 $Hz$), this implies that the modified current source is doubled compared to the initial current source [\cite{Grcev}]. Thus if the conductor is considered to be filiform and if again the fundamental solution (\ref{fundsol2}) is used, for a 3 phase system the total magnetic vector potential at distance $|\vec{r}_f|$ per phase, seen from the power line to the buried pipeline, becomes
\begin{equation}
\vec{A}=\frac{\mu}{2\pi}\sum^{3}_{f=1}\underline{I}_{s,f}\int_C \frac{e^{-\beta |\vec{r}_f|\left(1+i\right)}}{4\pi |\vec{r}_f|}\vec{1}_{s,f} d \Gamma,
\label{veca2}
\end{equation}
where the vector $\vec{1}_{s,f}$ is the unit vector in the direction of the power line per phase and $\underline{I}_{s,f}$ is the phase current and normally shifted 120 degrees per phase.
\\ \\
Using (\ref{veca2}) the contribution of each point of the power line to the buried pipeline of the magnetic vector potential can be computed. This is executed by discretization of the power line into $M$ elements of length $\ell$ whereby each element contributes to the total vector potential in each node of the discretized buried pipeline.
\\ \\
Note that in the general case the number of elements for discretization of the power line does not need to be equal to the number of elements for the discretization of the buried pipeline nor these elements need to be equidistant.
\\ \\
The complex magnetic vector potential $\vec{A}$ computed in each point through equation (\ref{veca2}) is further split per element $j$ into its axial $A_{z_j}$ respectively radial $A_{\rho_j}$ component by taking into account the angle between the pipeline and the power line. If $\vec{1}_{p_j}$ is the direction vector of the pipeline in element $j$ and $\vec{1}_{HV_i}$ is that of the $i$-th element of the power line from which the contribution is calculated, the partition is [\cite{Bortels}]
\begin{subequations}
\begin{equation}
A_{z_j}=|\vec{A}|\left( \vec{1}_{p_j}\cdot\vec{1}_{HV_i}\right ),
\end{equation}
\begin{equation}
A_{\rho_j}=\sqrt{|\vec{A}|^2-A^2_{z_j}},
\end{equation}
\label{veca4}
\end{subequations}
\\ \\
Hence, successively computing (\ref{veca2}), (\ref{veca4}a) and (\ref{veca4}b) considering the 3 phases yields the applied source conditions of the total system matrix.
\subsection{The total system matrix}
Combining (\ref{tsm1}), (20) and (23) yields the total system matrix
\\
\begin{align}
\begin{bmatrix}
\doubleunderline{S}^{p} & \doubleunderline{K}^{p} & 0 & 0 \\
\multicolumn{2}{c}{\multirow{2}{*}{$\underline{H}^{c}$}} & \multicolumn{2}{c}{\multirow{2}{*}{$\doubleunderline{G}^{c}/(\sigma_c+i\omega\epsilon_c)$}}\\ \\
0 & 0 & \doubleunderline{H}^{s} & \doubleunderline{G}^{s}/\sigma_s
\end{bmatrix}
\left \{
\begin{array}{c} \underline{\Phi} \\  \underline{J}_\rho \\ \underline{V} \\ \underline{Q}_\rho\end{array} \right \}
\nonumber
\\
 =
\begin{bmatrix}
\doubleunderline{A}^{p} & 0 \\
\multicolumn{2}{c}{\multirow{2}{*}{$0$}} \\ \\
0 &  i\omega \doubleunderline{G}^{s} \\ 
\end{bmatrix}
\left \{
\begin{array}{c} \underline{A}_z \\ 0 \\ \underline{A}_\rho  \end{array} \right \}.
\end{align}
By following the procedure described in section \ref{mvp} the applied boundary conditions in the RHS of the total system matrix is calculated resulting in a solvable system of the form
\begin{equation}
\doubleunderline{A}\ \underline{X}=\doubleunderline{B}.
\end{equation}
The one dimensional column $\underline{X}$ represents the unknowns i.e. the voltage $\Phi$ in the pipe, the voltage $V$ in the earth at the coating surface, the inner radial current density $J_\rho$ and the outer radial current density $Q_\rho$. Due to the use of (\ref{eq24}) the potential at infinity is fixed at $0V$.
\section{Validation of the model}
\subsection{Electromagnetic parameters}
The total system matrix is numerically implemented in a $C$-environment of XCode Version 6.3 (6D570) through which any arbitrary configuration of a buried pipeline in the presence of a high voltage power line can be calculated. The used electromagnetic simulation parameters are listed in table \ref{tab:2}.
\begin{table}[h]
\caption{Electromagnetic simulation parameters.}
\centering
 \small 
\begin{tabular}{llr}
\toprule
Pipeline outer radius $R_p$ & \quad & $0.25 m$\\
Wall thickness & \quad & $2 cm$\\
Steel pipe conductivity $\sigma_p$ & \quad & $5.88$ $10^6 (\Omega m)^{-1}$\\
\midrule
Soil conductivity & \quad & $0.01 (\Omega m)^{-1}$\\
\midrule
Coating conductivity $\sigma_c$ & \quad & $4$ $10^{-6} (\Omega m)^{-1}$ \\ 
Coating thickness $\delta_c$ & \quad & $4 mm$ \\
Relative permittivity of the coating $\epsilon_{r,c}$ & \quad & 5 \\
\midrule
Magnetic permeability of free space & \quad & $4\pi 10^{-7}$ $N / A^{2}$ \\
\midrule
HV-line type & \quad & SH345kV\_NS \\
Phase current $I_{s,f}$ & \quad & $500 A$\\
Phase 1 & \quad & 0 $rad$\\
Phase 2 & \quad & $2\pi/3$ $rad$\\
Phase 3 & \quad & $-2\pi/3$ $rad$\\
Frequency & \quad & $50Hz$\\
\midrule
\bottomrule
\end{tabular}
\label{tab:2}
\end{table}
\subsection{Overview of the configurations}
Two specific configurations are simulated. The first configuration is one of a pipeline that parallels a high voltage line for a few kilometers and ends at both ends. The second configuration is a real life situation where a pipeline approaches a high voltage line at a certain angle, runs parallel with it for a few kilometers and diverges away from it. The coordinates of both configurations as well as that of the high voltage line are listed in table \ref{chap10_tab4}.
\begin{table}[h]
\caption{Coordinates of the power line and buried pipeline.}
\centering
 \small 
\begin{tabular}{llll}
\hline \hline
&  x [km] & y [km] & z[m] \\
\hline
HV-line \\
\quad point 1 & -1 & 0 & 18,29 \\
\quad point 2 & 12 & 0 & 18,29 \\
\hline
Pipeline (configuration 1) \\
\quad point 1 & 4.5 & $0.025$ & -5\\
\quad point 2 & 7.5 & $0.025$ & -5\\
\quad Number of elements & & & 45\\
\quad Length of a pipe element & & & $66.7$m\\
\hline
Pipeline (configuration 2) \\
\quad point 1 & 0 & 2	& -5 \\
\quad point 2 & 2 & 1.025 & -5 \\
\quad point 3 & 3.1 & 0.495 & -5 \\
\quad point 4 & 3.75 & 0.18 & -5 \\
\quad point 5 & 4 & 0.057 & -5 \\
\quad point 6 & 5 & 0.057 & -5 \\
\quad point 7 & 6 & 0.057 & -5 \\
\quad point 8 & 7 & 0.057 & -5 \\
\quad point 9 & 7.057 & 0.029 & -5 \\
\quad point 10 & 7.094 & 0.010 & -5 \\
\quad point 11 & 7.114 & -0.010 & -5 \\
\quad point 12 &7.17 & -0.020 & -5 \\
\quad point 13 & 7.28 & -0.080 & -5 \\
\quad point 14 & 7.556 & -0.230 & -5 \\
\quad point 15 & 7.9 & -0.400 & -5 \\
\quad point 16 & 8 & -0.450 & -5 \\
\quad point 17 & 8.844 & -0.872 & -5 \\
\quad point 18 & 9.79 & -1.345 &	-5 \\
\quad point 19 & 10 &	-1.450 & -5 \\
\quad point 20 & 11 &	-1.950 & -5 \\
\quad point 21 & 12 &	-2.450 & -5 \\
\quad Number of elements & & & 261\\
\quad Length of a pipe element & & & $51.2$m\\
\hline \hline
\end{tabular}
\label{chap10_tab4}
\end{table}
\subsection{Simulation Results}
\begin{figure}
  \centering
  \includegraphics[width=0.9\linewidth]{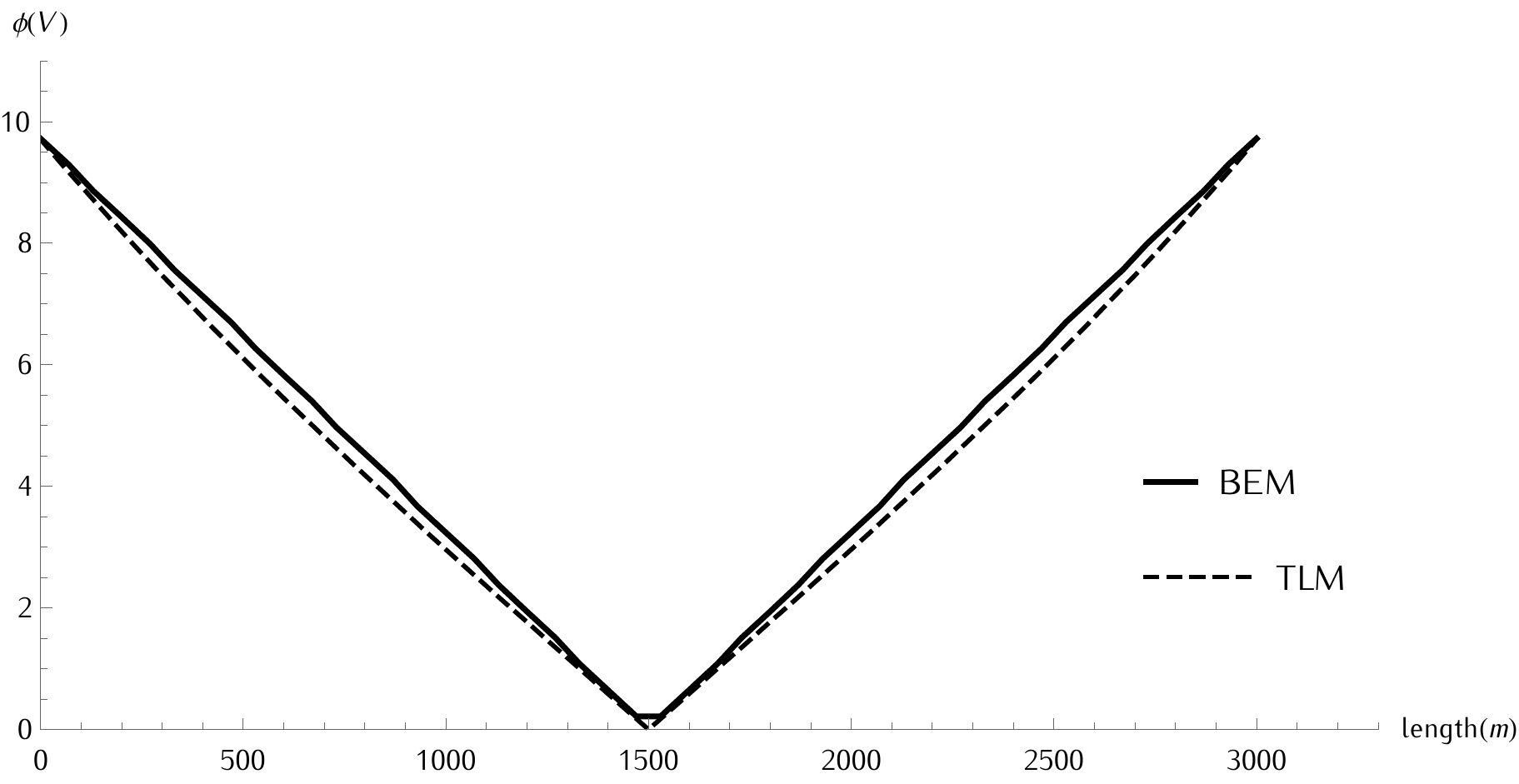}
  \caption{BEM model with coating layer: simuation results for configuration 1.}
\label{sim_conf1}
\end{figure}

\begin{figure}
  \centering
  \includegraphics[width=0.9\linewidth]{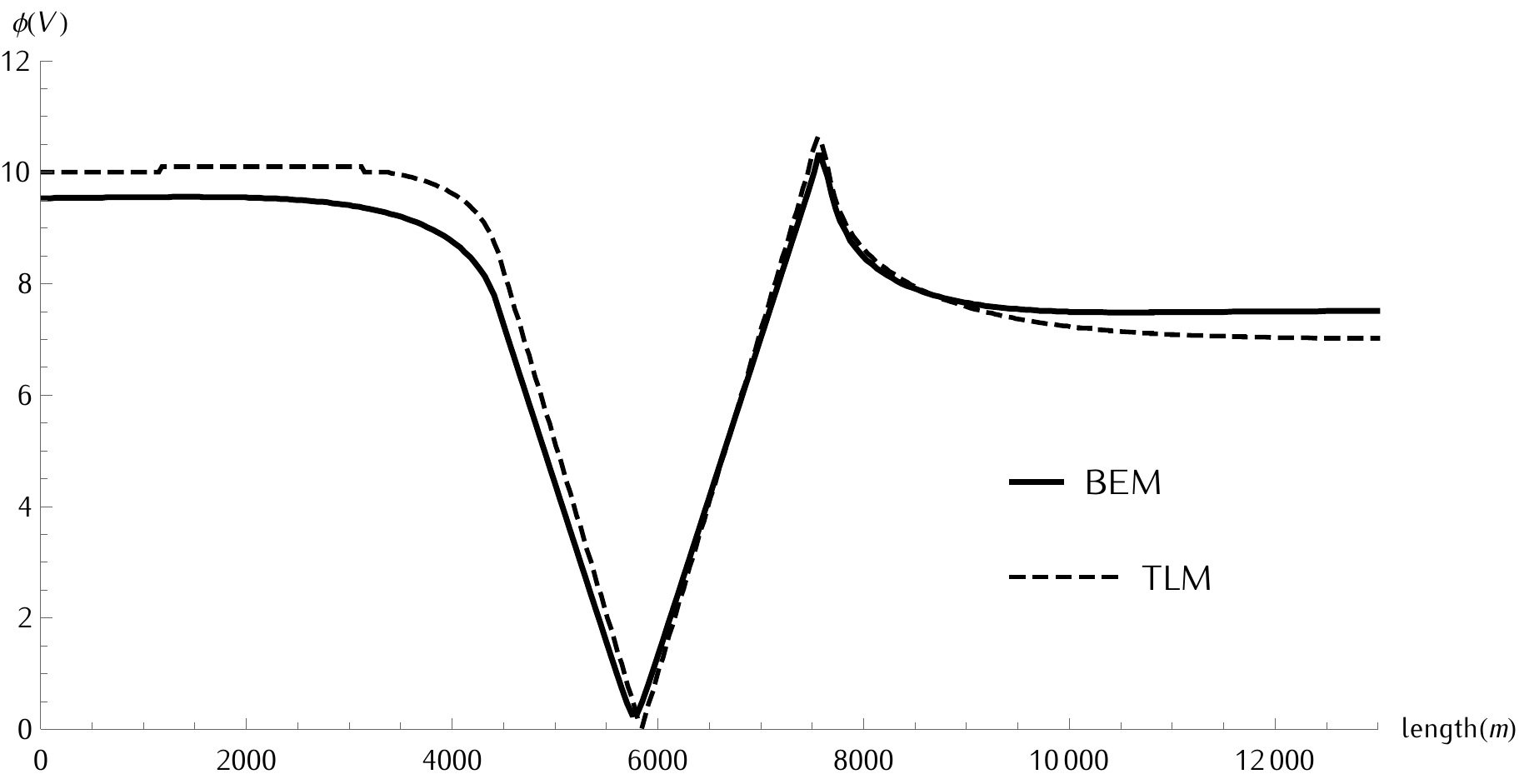}
  \caption{BEM model with coating layer: simuation results for configuration 2.}
\label{sim_conf2}
\end{figure}

In order to verify the model, the results are compared with those of a transmission line model discussed in \cite{Bortels}. Since this latter model already has proven its validity, this comparison is justified. Furthermore, it is also observed by the authors that the coating layer could not be modelled as an impedance between the internal problem using the FEM and the external problem using the BEM, since this approach would assume that $J_{\rho_i}=Q_{\rho_i}$ in each node. It was observed after numerous calculations that such an assumption in this application leads to numerical errors.
\\ \\
The simulation results of the first configuration together with those of the transmission line model are presented in figure \ref{sim_conf1}, while those of the second configuration are presented in figure \ref{sim_conf2}.
\\
\\
Observe that analytical solutions do not exist. Hence, both models, the transmission line model and the proposed one in this paper are approximations. However, since the transmission line model demands characteristic impedances at both ends of the configuration in order to simulate a pipeline that diverges away from a power line, it can be argued that the proposed model contains a better approach, since this condition is implemented in an intrinsic way.
\\ \\
It is observed that the simulation results of the proposed model using a fundamental solution which takes into account the capacitive effect of the considered medium are in good agreement with those of a transmission line model. This good agreement is valid for a simple as well as a complex configuration. Yet, the proposed method includes all different electromagnetic characteristics in a directly implemented way including the coating thickness and conditions at both ends of a pipeline.
\section{Conclusion}
In this paper a generalized fundamental solution when using the boundary element method for electromagnetic field problems is proposed. This generalized fundamental solution incorporates the capacitive effect when dealing with media wherein this effect is not negligible, such as coating layers.
\\ \\
Using this generalized fundamental solution, a model was developed which combined a finite element method with the boundary element method of a buried coated pipeline in close proximity to a high voltage line, where as source conditions the magnetic vector potential originating from this high voltage line is calculated.
\\ \\
The model is implemented by the authors in a C-environment, and any configuration can be calculated. Two configurations were implemented and the results were compared with the results of a transmission line model.
\\ \\
There is a slight difference between the two models that we assign to unavoidable approximations of both methods. This requires additional research and eventually comparison with field measurements. One can conclude that the proposed model with the generalized fundamental solution using the boundary element method is valid and can be used for industrial applications.
\newpage
\section*{Bibliography}
\bibliographystyle{elsarticle-harv} 

\begin{thebibliography}{17}
\expandafter\ifx\csname natexlab\endcsname\relax\def\natexlab#1{#1}\fi
\expandafter\ifx\csname url\endcsname\relax
  \def\url#1{\texttt{#1}}\fi
\expandafter\ifx\csname urlprefix\endcsname\relax\def\urlprefix{URL }\fi

\bibitem[{Balanis(2012)}]{Balanis}
Balanis, C., 2012. Advanced Engineering Electromagnetics, 2nd Edition. John
  Wiley \& Sons, Inc., USA.

\bibitem[{Bortels(2006)}]{Bortels}
Bortels, L., D. J. M. C. . T.~V., 2006. A general applicable model for ac
  predictive and mitigation techniques for pipeline networks influenced by hv
  power lines. IEEE Transactions on Power Delivery 21~(1), 210--217.

\bibitem[{Brichau(1994)}]{Brichau}
Brichau F., 1994. A Mathematical Model for the Cathodic Protection of
  Underground Pipelines including Stray Currents.
  Ph.D. Thesis, Vrije Universiteit Brussel, Belgium.

\bibitem[{Brebbia(1989)}]{Brebbia}
Brebbia, C.A. \&~Dominguez, J., 1989. Boundary Elements: An Introductory
  Course. Computational Mechanics Publications, McGraw-Hill Book Company, New
  York, USA.

\bibitem[{Carson(1926)}]{Carson}
Carson, J., 1926. Wave propagation in overhead wires with ground return. Bell
  System Technical Journal Vol.5, 339--35.

\bibitem[{CIGRE(1995)}]{CIGRE}
CIGRE, Working Group 36.02 (1995). Guide to the influence of High Voltage AC Power Systems on Metallic Pipelines.

\bibitem[{Dawalibi(1986)}]{Dawalibi}
Dawalibi, F.P. \&~Pinhi, A., 1986. Computerized analysis of power systems and
  pipelines proximity effects. IEEE Transactions on Power Delivery 4~(2),
  40--48.

\bibitem[{Dawalibi(1989)}]{Dawalibi2}
Dawalibi, F.P. \&~Southey, R., 1989. Analysis of electrical interference from
  power lines to gas pipelines, part i: Computation methods. IEEE Transactions
  on Power Delivery 4~(3), 1840--1846.

\bibitem[{Gary(1985)}]{Gary}
Gary, C., 1985. Nouvelle methode de calcul des inductances propres et mutuelles
  des lignes en presence du sol. CIGRE Symposium 06-85.

\bibitem[{Griffiths(1999)}]{Griffiths}
Griffiths, D., 1999. Introduction to Electrodynamics: Third Edition. Prentice
  Hall International, Inc., Upper Saddle River, New Jersey, USA.

\bibitem[{L.D.(1996)}]{Grcev}
L.D., G., 1996. Computer analysis of transient voltages in large grounding
  system. IEEE Transactions on Power Delivery 11~(2), 815--823.

\bibitem[{Munteanu(2012)}]{Munteanu}
Munteanu, C., M. G. P. M. T. V. P. I. G. L. . R.~C., June 2012. Electromagnetic
  field model for numerical computation of voltages induced on buried pipelines
  by high voltage overhead power lines. The European Physical Journal Applied
  Physics 58~(3), 30902 (9).

\bibitem[{Reitz(1979)}]{Reitz}
Reitz, J.R., M. F. . C.~R., 1979. Foundation of Electromagnetic Theory: Third
  Edition. Jaddison-Wesley Publishing Company, Inc., USA.

\bibitem[{Sadiku(2001)}]{Sadiku}
Sadiku, M., 2001. Numerical Techniques in Electromagnetics, Second Edition. CRC
  Press LLC, Florida, USA.

\bibitem[{Schoonjans(2017)}]{Schoonjans}
Schoonjans, B. \&~Deconinck, J., 2017. Calculation of HVAC inductive coupling
  using a generalized BEM for helmholtz equations in unbounded regions.
  International Journal of Electrical Power \& Energy Systems 84 (2017) 242-251.

\bibitem[{Shen(1995)}]{Shen}
Shen, J., 1995. Topics in Engineering Vol.24: Computational Electromagnetics
  using Boundary Elements - Advances in Modelling Eddy Currents. Computional
  Mechanics Publications, Southampton UK and Boston, USA.

\bibitem[{Stratton(2007)}]{Stratton}
Stratton, J., 2007. A Classic Reissue: Electromagnetic Theory. IEE Press Series
  on Electromagnetic Wave Theory, Wiley-Interscience.

\bibitem[{Sunde(1966)}]{Sunde}
Sunde, E., 1966. Earth Conduction Effects in Transmission Systems, Chapter 5.
  Van Nostrand, New York, USA.

\bibitem[{Sykulski(1995)}]{Sykulski}
Sykulski, J., 1995. Computational Magnetics. James \& James (Science
  Publishers) Ltd and Chapman \& Hall, London, UK.

\end{thebibliography}

\newpage
\section*{Nomenclature}
\begin{tabular}{llr}
\toprule
$\vec{A},A_\rho,A_z$ & Magnetic vector potential & $[Vs/m]$\\
$\doubleunderline{A}^{p}$ & FEM matrix related with & \\
 & \quad the axial magnetic vector potential & \\
$\vec{B}$ & Magnetic flux density & $[T]$\\
$c_i$ & BEM correction factor & \\
$\vec{E}$ & Electric field & $[V/m]$\\
$E$ & Electromotive force induced & \\
& \quad per unit length & $[V/m]$\\
$f$ & Frequency & $[Hz]$\\
$\mathcal{G}$ & Fundamental solution & \\
$\doubleunderline{G}^{p}$ & BEM matrix related with & \\
 & \quad the current density and & \\
 & \quad the radial magnetic vector potential & \\
$\doubleunderline{H}^{p}$ & BEM matrix related the potential & \\
$i$ & Imaginary unit & \\
$\vec{J},J_\rho$ & Current density & $[A/m^2]$\\
$k$ & Surface of revolution & $[m^2]$\\
$\doubleunderline{K}^{p}$ & FEM matrix related with & \\
 & \quad the current densitiy & \\ 
$\tilde{k}$ & Complex wave number & \\
$l$ & Length of pipe element & $[m]$\\
$\vec{n}$ & Normal unit vector & \\
$N$ & Number of elements & \\
$\vec{r}$ & Vector to source point & \\
$\vec{r}'$ & Vector to image point & \\
$\doubleunderline{S}^{p}$ & FEM matrix related with & \\
 & \quad the potential & \\
$S_{p}$ & Axial conductivity of the pipe & $[1/ \Omega m]$\\
\\
\\
$\alpha$ & Attenuation transmission coefficient & $[Np/m]$\\
$\beta$ & Phase transmission coefficient & $[rad/m]$\\
$\delta$ & Skin depth & $[m]$\\
$\epsilon_0$ & Electric constant & $[F/m]$\\
$\epsilon$ & Relative permittivity & \\
$\mu_r$ & Relative permeability &\\
$\nu$ & Coefficient expressing a scaled radius & \\
$\sigma_c$ & Conductor electrical conductivity & $[1/\Omega m]$\\
$\sigma_{s}$ & Soil electrical conductivity & $[1/\Omega m]$\\
$\Phi$ & Electrical potential & $[V]$\\
$\omega$ & Angular frequency & $[rad/s]$\\
$| \ldots |$ & Magnitude of vector\\
\bottomrule
\end{tabular}
\end{document}